\documentclass{elsart}

\usepackage{graphicx}

\usepackage{amssymb}

\begin{document}

\begin{frontmatter}




\title{Shape of ammonium chloride dendrite tips at small supersaturation}
\author{Andrew Dougherty} and
\ead{doughera@lafayette.edu}
\ead[url]{http://www.lafayette.edu/\%7Edoughera/}
\author{Mayank Lahiri}
\address{Dept.~of Physics, Lafayette College, Easton PA 18042}

\begin{abstract}
We report detailed shape measurements of the tips of
three-dimensional NH$_4$Cl dendrites grown from
supersaturated aqueous solution.
For growth at small supersaturation, we compare
two models for the tip shape: parabolic with a fourth-order
correction and a power law.  Neither is ideal, but the fourth-order fit
appears to provide the most robust description of both the tip
shape and position for this material.
For that fit, the magnitude of the fourth-order coefficient is about 
half of the theoretically expected value.

\end{abstract}

\begin{keyword}
A1. Dendrites \sep A2. Growth from solution \sep A1. Crystal morphology
\PACS 68.70.+w \sep 81.10.Dn \sep 64.70.Dv
\end{keyword}
\end{frontmatter}

\section{Introduction}

Dendrites are a commonly observed microstructural form
resulting from the diffusion-limited solidification of non-faceting
materials, and they continue to be interesting for both practical
and aesthetic reasons.  Practically, an improved understanding of
dendritic microstructures may enhance the ability to predict and
control material properties.  Aesthetically, they are an intriguing
example of pattern formation under non-linear and non-equilibrium
conditions.\cite{Trivedi00,Glicksman-rev,Kessler88a,Langer80}

Two of the most basic experimental characterizations of a growing dendrite
are the tip size and growth speed.  Although there is already
considerable data
available, recent advances in both theory and experimental technique
have made more precise comparisons between them possible.
Some comparisons have already
been made for some pure materials\cite{Glicksman95,Bilgram96c}, but it
is important that they also be made for as broad a range of systems
as is reasonable, in order to clarify the roles of various effects
such as solution \textit{vs}.\ thermal growth, and different values of
crystalline anisotropy.

In the absence of surface tension, one solution to the diffusive
growth problem is a parabolic dendrite with radius of
curvature $\rho$ propagating at constant speed $v$.
However, the presence of surface tension, and
the instabilities that lead to sidebranching, complicate the problem
considerably.  Indeed even the most basic issues about the precise tip
shape and whether a dendrite actually grows with a constant velocity
are still areas of active research
\cite{Glicksman95,Bilgram96c,Bilgram96b,Glicksman99b,Glicksman99c,Glicksman99d,Karma00,McFadden00,Glicksman02a,Bilgram96d}.

Addressing those issues requires unambiguous ways to identify both the tip
size and position.  In this paper, we consider two different models for
the dendrite tip shape and evaluate their use to characterize the growth
of NH$_4$Cl dendrites at small supersaturations.  A similar exploration
of different tip shape models for dendrites resulting from phase-field
simulations was reported by Karma, Lee, and Plapp\cite{Karma00}, but
it is important to investigate how well the different models work for
different materials under actual observation conditions.

\section{Background}

\begin{figure}
\includegraphics[keepaspectratio=true, width=\columnwidth]{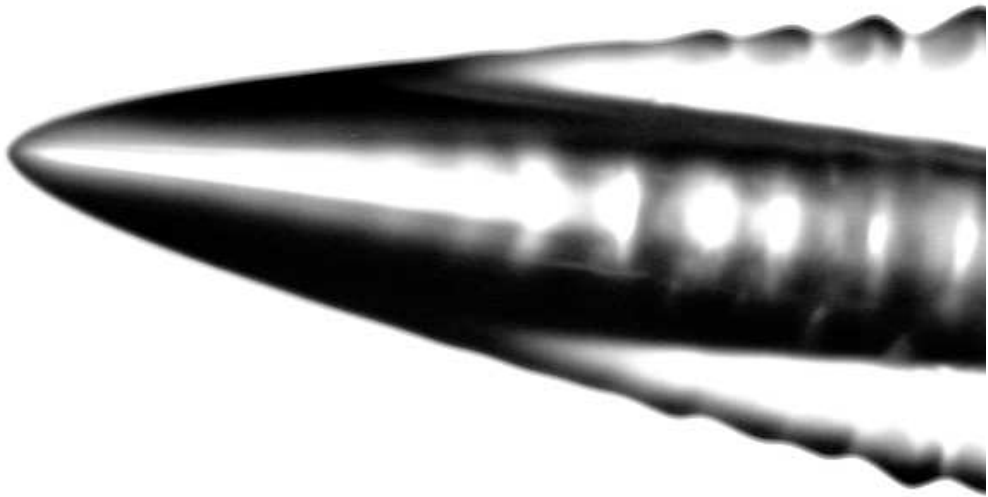}
\caption{\label{photo1}Dendrite of NH$_4$Cl.  The picture is
approximately 400 $\mu$m across.}
\end{figure}

A typical dendrite of NH$_4$Cl grown in this study is shown in
Fig.~\ref{photo1}.
Since NH$_4$Cl has cubic symmetry, four sets of sidebranches
grow approximately perpendicular to the main dendrite stem.
In Fig.~\ref{photo1}, two sets of sidebranches are visible in the plane
of the image; two additional branches are growing perpendicular to the
plane of the image.

\begin{figure}
\includegraphics[keepaspectratio=true, width=\columnwidth]{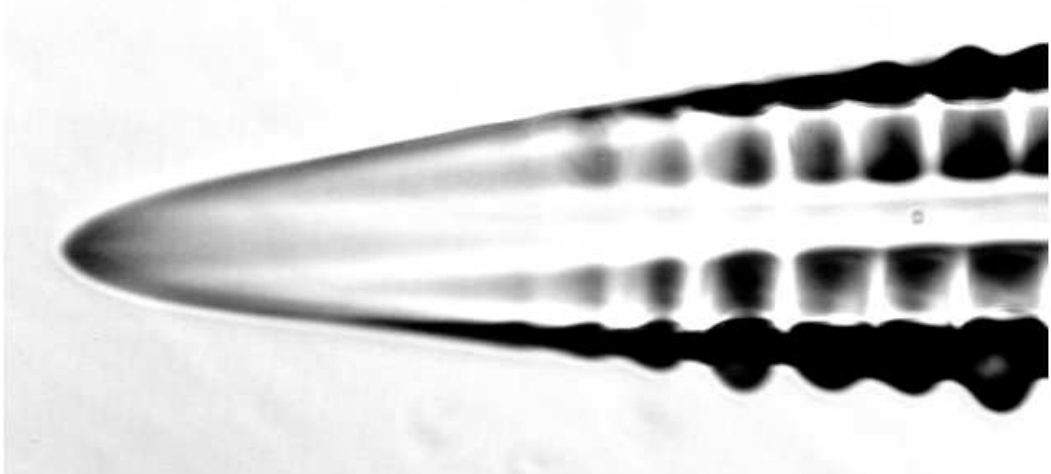}
\caption{\label{photo2}Dendrite of NH$_4$Cl with $\phi \approx 45^\circ$.
The scale is the same as in Fig~\ref{photo1}.}
\end{figure}

The coordinate system used in this work is defined as follows.
The main dendrite stem is used to define the $z$ axis.  The growth
direction is taken as the negative-$z$ direction, so the solid crystal
lies along the positive $z$ axis.
The $x$ axis is defined as the
direction in the plane of the image perpendicular to $z$.  Lastly,
$\phi$ is defined to be the rotation angle of the crystal
around the $z$-axis.  A
dendrite with $\phi \approx 45^\circ$ is shown in
Fig.~\ref{photo2}.

\subsection{Parabolic fit with fourth-order correction}

For diffusion-controlled growth in the absence of surface tension,
the Ivantsov solution is a paraboloid of revolution of radius
$\rho$ growing at speed $v$.
Once anisotropic surface tension is included,
microscopic solvability\cite{Kessler88a,Kessler88b} predicts
both the tip size $\rho$ and speed $v$ depend on the crystalline anisotropy
$\epsilon_4$.
For a fourfold-symmetric crystal such as NH$_4$Cl,
that anisotropy can be expressed in spherical coordinates
as\cite{McFadden00}
\begin{equation}
\gamma (\Theta, \Phi)/\gamma_0 =
1 + \epsilon_4 [ 4 \cos ^4 \Theta + (3 + \cos 4 \Phi) \sin ^4 \Theta]
\label{eq:eps4}
\end{equation}
where $\gamma$ is the surface free energy, and $\Theta$ and $\Phi$
are the usual spherical angles.

Although the presence of anisotropic surface tension is critical for
the development of the dendritic structure, the overall magnitude of the
surface tension is small, so the deviations of the tip shape from
parabolic might be expected to be rather small as well.
Many experiments, including the landmark experiments of Huang and
Glicksman\cite{Glicksman81a}, confirm that a parabola is indeed
a reasonably good approximation to the tip shape.  Experiments on
NH$_4$Br also showed that a parabola is a reasonably good approximation,
at least relatively close to the tip\cite{Dougherty88}.

In the limit of small fourfold-anisotropy, Ben
Amar and Brener\cite{BenAmar93} found that
the lowest order correction to the parabolic shape is
a fourth-order term proportional to
$\cos 4 \phi$, where $\phi$ is the rotation angle around the $z$ axis.
Thus, at least close to the tip, the tip shape
could be reasonably well described by
\begin{equation}
z = z_{tip} + \frac{(x-x_{tip})^2}{2\rho} -
A_4 \cos 4 \phi \frac{(x-x_{tip})^4}{\rho^3} ~,
\label{eq:fourth}
\end{equation}
where $(x_{tip}, z_{tip})$ is the location of the tip,
$\rho$ is the radius of curvature at the tip,
and $A_4 = 1/96$,
independent of anisotropy strength\cite{Brener95}.
Using a somewhat different approach, McFadden, Coriell, and
Sekerka\cite{McFadden00} found that, to second order in
$\epsilon_4$,
$A_4 = \epsilon_4 + 12 \epsilon_4 ^2$, at least very close to the tip.
Using an estimate of $\epsilon_4 \approx 0.016$ (the value reported for
NH$_4$Br\cite{Dougherty88}) this corresponds to $A_4 \approx 0.019$.

Tip shapes consistent with this model were found by LaCombe and
coworkers.\cite{Glicksman95,Glicksman99b} They studied succinonitrile
dendrite tips under a variety of three-dimensional crystal orientations,
evaluated both second order and fourth-order polynomial fits,
and concluded that the fourth-order fit worked significantly
better.\cite{Glicksman95,Glicksman99b}

\subsection{Power-law}

Further back from the tip, the crystalline anisotropy tends to
concentrate material into four ``fins'' such that the shape is no longer
well-described by Eq.~(\ref{eq:fourth}).  The width of the fins
is predicted\cite{Brener93} to scale as $(z - z_{tip}) ^ {3/5}$.
Scaling consistent with this prediction was observed in the
average shapes of NH$_4$Cl and pivalic acid dendrites grown from
solution\cite{Dougherty92,Dougherty94}.

Although this power-law scaling was originally proposed to
describe the shape of the crystal in the region {\em behind} the
tip, Bisang and Bilgram found that for xenon dendrites with $\phi
\approx 0$, this power law was a good fit even quite close to the
tip.\cite{Bilgram96d,Bilgram95b}.  Hence the power law offers another
way to characterize the tip size and location.

In this model, we describe the tip shape by
\begin{equation}
z = z_{tip} + \frac{|x-x_{tip}|^{5/3}}{(2 \rho)^{2/3}},
\label{eq:power}
\end{equation}
where we have included the factor of 2 by analogy with
Eq.~\ref{eq:fourth}.
The parameter $\rho$ still sets a length scale for the dendritic
structure,
but the curvature at the tip is no longer defined.

\subsection{Experimental Considerations and Model Limitations}

Each of these models has different limitations.  From the theoretical
perspective, the fourth order fit is only appropriate very close
to the tip, well before sidebranches become significant.  Hence in
order to avoid contamination from sidebranches, only data with $(z -
z_{tip}) < z_{max}$ should be used in the fit.  Since sidebranching
activity is detectable even close to the tip (at least for NH$_4$Br
dendrites\cite{Dougherty87}), $z_{max}$ should not be made too large.
On the other hand, $z_{max}$ should not be made too small since the the
region around the tip contains the sharpest curvatures and the largest
concentration gradients, and hence is the the most subject to optical
distortions\cite{Glicksman95,Dougherty88}.  Considering both issues,
Dougherty and Gollub\cite{Dougherty88} suggested $z_{max} = 3 \rho$
as a compromise for parabolic fits.

In contrast, the power-law fit is only appropriate further behind the
tip, so its usefulness for describing the tip size and position must
be explicitly tested.  However, since it is not necessarily
constrained to as small
$z_{max}$, the fit can include data points less contaminated by optical
distortions near the tip.  On balance, for xenon dendrites with $\phi
\approx 0$, Bisang and Bilgram found that this power law provided a
reasonable fit.

Both models are potentially sensitive to the choice of $z_{max}$ used
in the fitting procedure, though such dependence ought to be minimal if
an appropriate fitting function is used.  Singer and Bilgram discuss
a procedure to determine $\rho$ from polynomial fits in a way that is
somewhat less model dependent\cite{Bilgram04b}, but that approach did
not offer any significant advantage for this system.

Thus both the fourth-order and the power-law fit provide reasonable fits,
at least in some cases, but a direct comparison of the two models for
the same material is required for an accurate assessment.

\section{Experiments}

The experiments were performed with aqueous solutions of ammonium chloride
with approximately 36\% NH$_4$Cl by weight.
The saturation temperature was approximately 65$^\circ$C.
The solution was placed in a $45 \times 12.5 \times 2$ mm glass cell
and sealed with a teflon stopper.  The cell was mounted in a massive
temperature-controlled copper block surrounded by an insulated
temperature-controlled aluminum block, and placed on the stage of
an Olympus BH-2 microscope.  The entire microscope was
enclosed in an insulating box.

The temperature of the outer aluminum block was controlled by an Omega
CN-9000 controller to approximately $\pm 1^\circ$C.  The temperature of
the inner copper block was controlled directly by computer.  A thermistor
in the block was connected via a Kiethley 2000 digital multi-meter to the
computer, where the resistance was converted to temperature.  The computer
then controlled the heater power supply using a software version of a
proportional-integral controller.  This allowed very flexible control over
not just the temperature, but also over any changes in the temperature,
such as those used to initiate growth.  The temperature of the sample
was stable to within approximately $\pm 5 \times 10^{-4\ \circ}$C.

A charged coupled device (CCD) camera was attached to the microscope
and images were acquired directly into the computer with a Data
Translation DT3155 frame grabber with a resolution of $640 \times 480$
pixels.  The ultimate resolution of the images was $0.63 \pm 0.01
\mu$m/pixel.  As a backup, images were also recorded onto video tape
for later use.

To obtain crystals, the
solution was heated to dissolve all the NH$_4$Cl, stirred to eliminate
concentration gradients, and then cooled to initiate growth.  Typically,
many crystals would nucleate.  An automated process was set up to acquire
images and then slowly adjust the temperature until all but the largest
crystal had dissolved.

This isolated crystal was allowed to stabilize for several days.
The temperature was then reduced by 0.77 $^\circ$C and the crystal was
allowed to grow.  The crystal was initially approximately spherical,
but due to the cubic symmetry of NH$_4$Cl, six dendrite tips would begin
to grow.  The tip with the most favorable orientation was followed,
and images were recorded at regular intervals.

The interface position was determined in the same manner as in
Ref.~\cite{Dougherty88}.  The intensity in the image was measured
on a line perpendicular to the interface.  Over the range of a few
pixels, the intensity changes rapidly from bright to dark.  Deeper inside
the crystal, the intensity begins to rise again.  (This corresponds
to the brighter areas inside the crystal in Fig.~\ref{photo1}.) In
the bright-to-dark transition region, we fit a straight line to the
intensity profile.  We define the interface as the location where the
fitted intensity is the average of the high value outside the crystal
and the low value just inside the crystal.

This fitting procedure interpolates intensity values and allows
a reproducible measure of the interface to better than one pixel
resolution.  It is also insensitive to absolute light intensity levels,
to variations in intensity across a single image, and to variations
in intensity \textit{inside} the crystal well away from the interface.
For the simple shapes considered here, this method is more robust and
requires less manual intervention than a more general contour-extraction
method, such as the one described by Singer and Bilgram for more complex
crystal shapes\cite{Bilgram04a}.

This method works best if the image is scanned along a line perpendicular
to the interface.  Since the position and orientation of the
interface are originally unknown, an iterative procedure is used
until subsequent iterations make no significant change to the fit.
Specifically, we start with an initial estimate of the size, location,
and orientation of the tip, and use that to scan the image to obtain a
list of interface positions up to a distance $z_{max}$ back from the tip
(where $z_{max}$ is some multiple of $\rho$).  We next rotate the data by
an angle $\theta$ in the $x-z$ plane and perform a non-linear regression
on Eq.~\ref{eq:fourth} or \ref{eq:power} to determine the best fit values
for $x_{tip}$, $z_{tip}$, $\rho$, and $A_4$ (if applicable), and the
corresponding $\chi^2$ value.  We then repeat with different $\theta$
values and minimize $\chi^2$ using Brent's algorithm.\cite{NumRecipes}
This completes one iteration of the fitting procedure.  We use this result
to rescan the image along lines perpendicular to the interface to obtain
a better estimate of the interface location and begin the next iteration.
The procedure usually converges fairly rapidly.  Even for a relatively
poor initial estimate, it typically takes fewer than 20 iterations.

There are some subtleties to the procedure worth noting.  First,
for a typical crystal in this work such as in Fig.~\ref{photo1}, only
about 120 data points are involved in the fit for $z_{max} = 6 \rho$.
For the fourth-order fit with five free parameters, there are often a
number of closely-spaced local minima in the $\chi^2$ surface, with tip
positions and radii varying by a few hundredths of a micrometer.  If the
iterative procedure enters a limit cycle instead of settling down to a
single final value, we select the element from that limit cycle with the
minimum $\chi^2$.  Second, it is worth noting that a generic fourth-order
polynomial fit is inappropriate for Eq.~\ref{eq:fourth}, since (after
rotation) there are only four parameters to fit: $x_{tip}$, $y_{tip}$,
$\rho$, and $A_4$.  Finally, we have no way to control or precisely
measure the orientation angle $\phi$ of the crystals in our system.

\section{Results}

\begin{figure}
\includegraphics{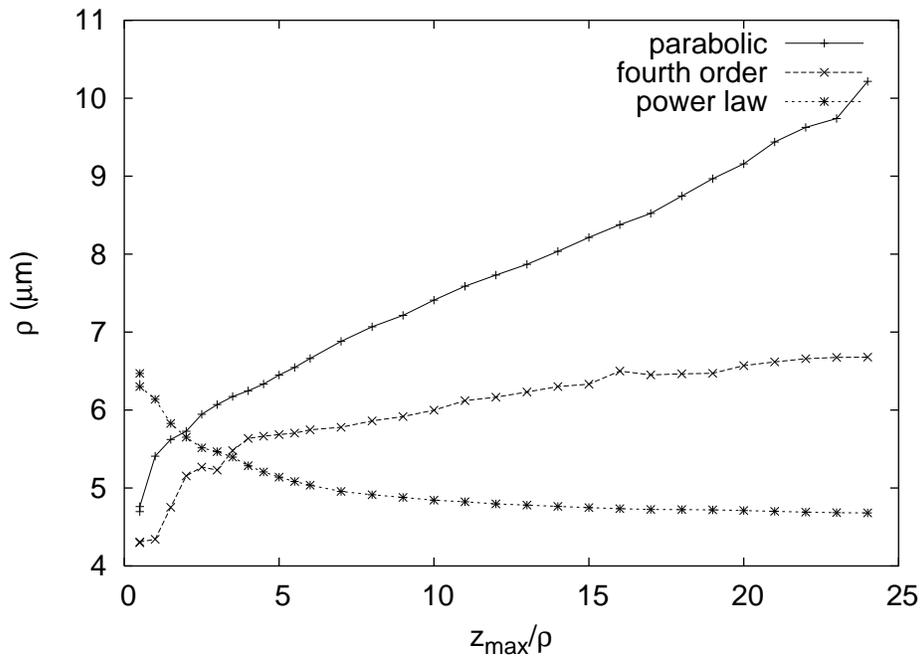}
\caption{\label{rho-z:0}Best fit value of $\rho$ as a function of
$z_{max}/\rho$ for the crystal shown in Fig.~\ref{photo1}.
The curves are for a parabolic fit ($+$), a fourth-order fit
($\times$), and a power-law fit ($*$).}
\end{figure}

\begin{figure}
\includegraphics{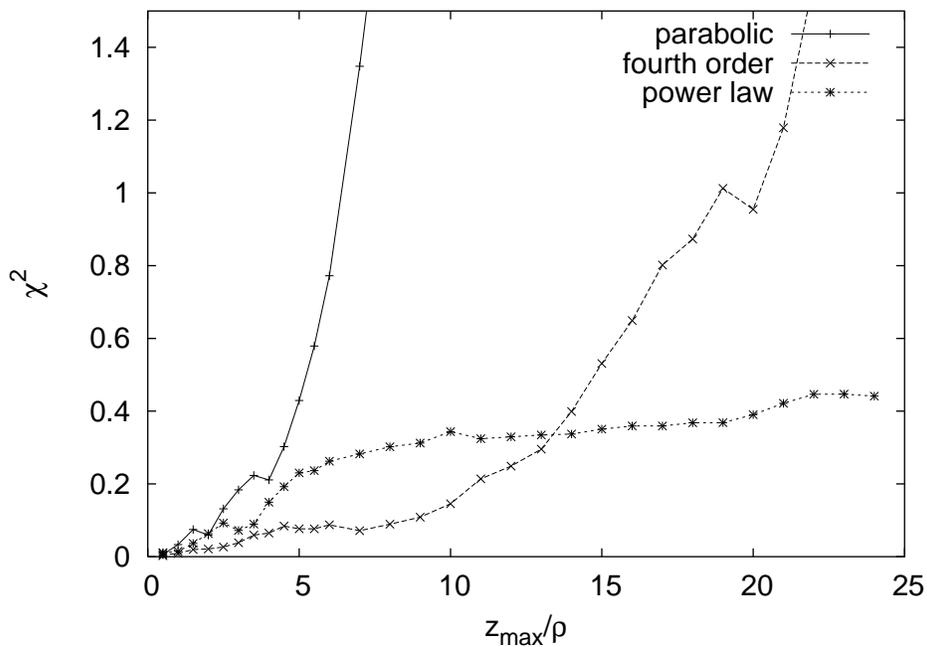}
\caption{\label{chisq-z:0}Values of $\chi^2$ for the best fit
as a function of $z_{max}/\rho$ for the crystal shown in Fig.~\ref{photo1}.
The curves are for a parabolic fit ($+$), a fourth-order fit
($\times$), and a power-law fit ($*$).}
\end{figure}

The best estimates of $\rho$ as a function of
$z_{max}/\rho$
for the crystal in Fig.~\ref{photo1}
are shown in Fig.~\ref{rho-z:0}.  We have included results for
the fourth-order and power-law fits as well as for a simple
parabola for comparison.
The corresponding $\chi^2$ values are shown in Fig.~\ref{chisq-z:0}.

None of the fits is robust very close to the tip, indicating that the
actual tip shape is not well-described by any of the candidate
functions.  The parabolic fit also gets rapidly worse for $z_{max}$
greater than about 5~$\rho$.  The fourth-order fit appears to have a
plateau between roughly 5 and 8$\rho$, but $\rho$ gradually increases
with $z_{max}$, and the $\chi^2$ value rapidly increases for $z_{max}
> 10 \rho$.  By contrast, the power law fit gives relatively stable
values at large $z_{max}$ for both $\rho$ and $\chi^2$.

\begin{figure}
\includegraphics{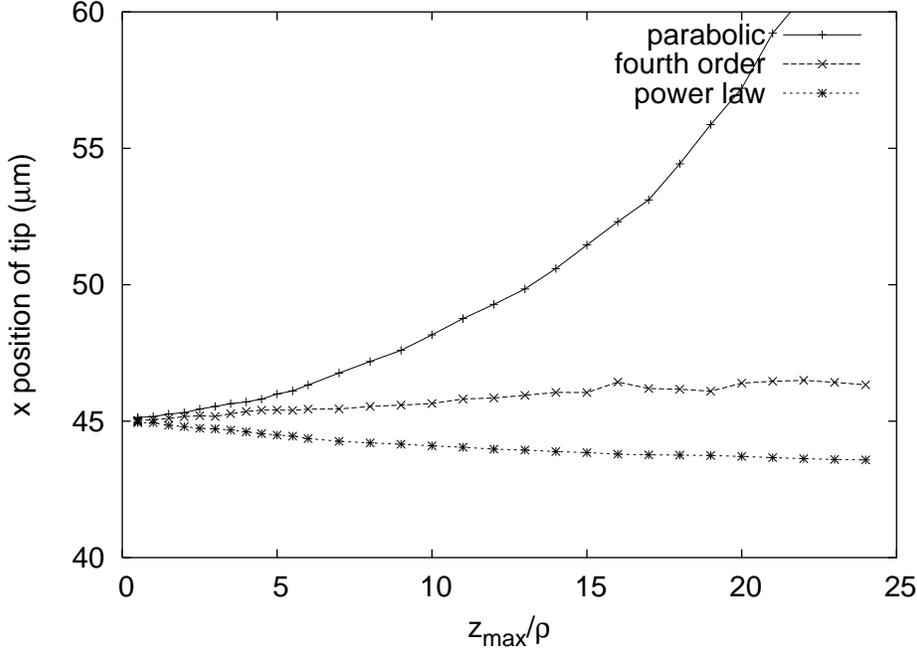}
\caption{\label{xtip-z:0}Best-fit values for the $x$-position of the
tip as a function of $z_{max}/\rho$ for the crystal shown in
Fig.~\ref{photo1}.  Symbols are as in the previous figures.}
\end{figure}

A second important consideration is the degree to which each fit
accurately describes the tip location.  This is illustrated in
Fig.~\ref{xtip-z:0}, which shows $x_{tip}$ for each fit.
Here again, the fourth-order fit is slightly more robust than the
power-law.

\begin{figure}
\includegraphics[keepaspectratio=true]{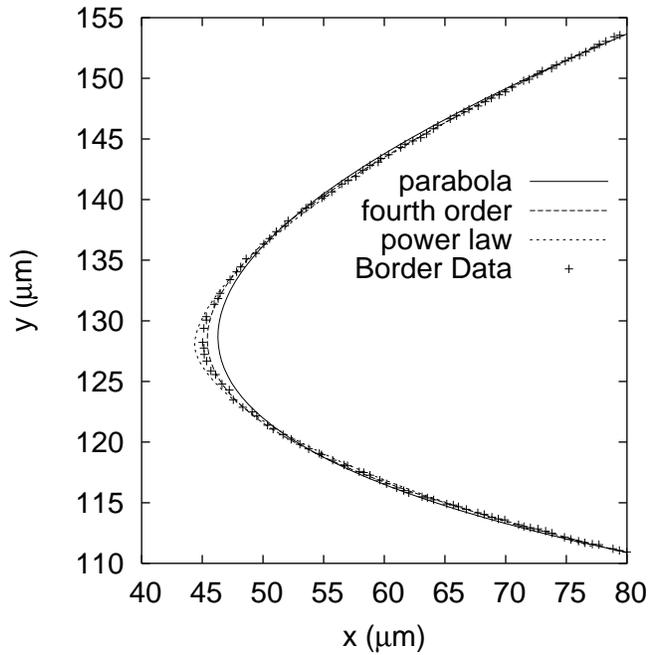}
\caption{\label{curves}Best-fit parabolic (solid), fourth-order
(dashed), and power-law (dotted)
curves for $z_{max} = 6 \rho$, along with the original border ($+$) for
the crystal shown in Fig.~\ref{photo1}.  Near the tip, the parabolic fit
is too far to the right, while the power law fit is too far to the
left.  The fourth-order fit matches the tip region fairly well.}
\end{figure}

Finally, Fig.~\ref{curves} shows the original data along with each
of the three model fits superposed for $z_{max} = 6 \rho$.  Of the
three fits, the fourth-order does the best job of capturing both the tip
location and shape.

For the value of $z_{max} = 6 \rho$, we find $A_4 = 0.004 \pm 0.001$,
which is similar to that measured by LaCombe {\it et al.} for
succinonitrile\cite{Glicksman95,Glicksman99b}, and to that obtained in
the simulations
by Karma, Lee, and Plapp\cite{Karma00}.
This value for $A_4$ is less than the value of
1/96 predicted by Brener\cite{Brener95}, and also significantly less
than the value of 0.019 estimated
by McFadden, Coriell, and Sekerka\cite{McFadden00}.
It is worth noting, however, that these predictions are only intended
to be valid close to the tip, where our fit is not robust.

\begin{figure}
\includegraphics{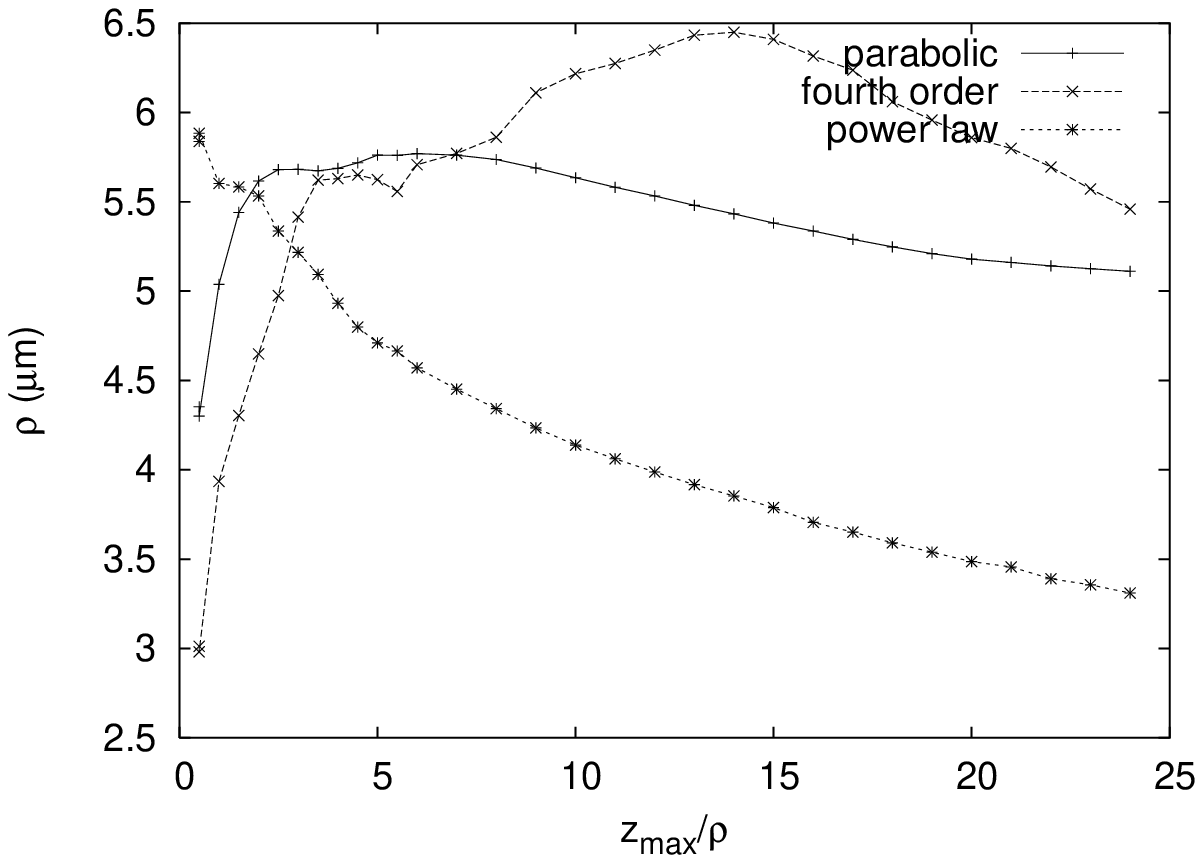}
\caption{\label{rho-z:45}Best fit value of $\rho$ as a function of
$z_{max}/\rho$ for the crystal shown in Fig.~\ref{photo2}.
The curves are for a parabolic fit ($+$), a fourth-order fit
($\times$), and a power-law fit ($*$).}
\end{figure}

\begin{figure}
\includegraphics{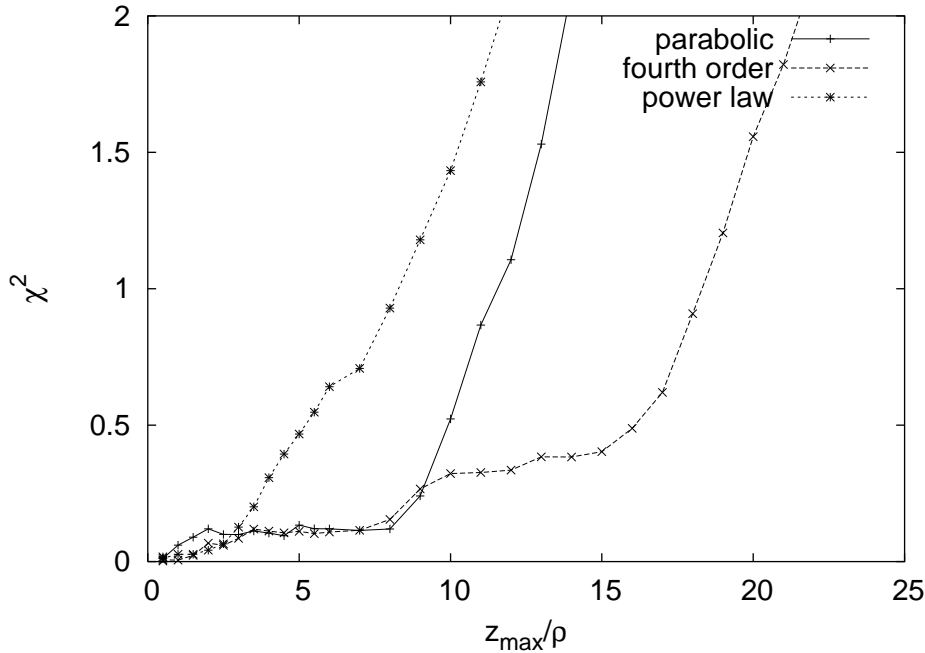}
\caption{\label{chisq-z:45}Values of $\chi^2$ for the best fit
as a function of $z_{max}/\rho$ for the crystal shown in Fig.~\ref{photo2}.
The curves are for a parabolic fit ($+$), a fourth-order fit
($\times$), and a power-law fit ($*$).}
\end{figure}

The results are slightly different for the crystal shown in Fig.~\ref{photo2},
which has $\phi \approx 45^{\circ}$.
The best estimates of $\rho$ as a function of
$z_{max}/\rho$ for the three fits
are shown in Fig.~\ref{rho-z:45},
and the corresponding $\chi^2$ values are shown in Fig.~\ref{chisq-z:45}.

Both the parabolic and fourth order models fit reasonably well for
$z_{max}$ between roughly $5$ and $8 \rho$.  Indeed within that
range, for the entire run from which Fig.~\ref{photo2} was taken, the
value for $A_4$ is $0.0007 \pm 0.0009$, consistent with zero.
In contrast, the power law is a poor fit.

One significant problem with this measurement is that the image
in Fig.~\ref{photo2} is a projection of the true three-dimensional
shape.  This is discussed in Ref.~\cite{Glicksman95,Glicksman99b}
and in considerably more detail in Ref.~\cite{Karma00}, but, in
general, our findings are consistent with those of LaCombe {\it et
al.}\cite{Glicksman95,Glicksman99b}.

\section{Conclusions}

We have considered two different models for the tip shape: parabolic with
a fourth-order correction, and power law.  For crystals oriented such
that $\phi \approx 0$, both give reasonable fits, though the fourth-order
fit is slightly better. For rotated crystals, such as those in
Fig.~\ref{photo2}, however, the fourth-order fit is significantly more
robust.

For the crystal in Fig.~\ref{photo1}, the coefficient of the fourth-order
term is $A_4 = 0.004 \pm 0.001$, significantly less than the theoretically
expected value.  These findings are consistent with those of LaCombe
{\it et al.}\cite{Glicksman95,Glicksman99b} for succinonitrile dendrites.

By contrast, the power-law fit was reasonably robust for
$\phi \approx 0$, in agreement with the results of Bisang and
Bilgram\cite{Bilgram96d,Bilgram95b} for xenon dendrites, but it did not
work as well for crystals with $\phi \approx 45^\circ$.  (One important
feature of the experiments in Refs.~\cite{Bilgram96d,Bilgram95b} was
the ability to control the viewing angle and hence $\phi$.)

One other
problem with the power law fit is that it does not accurately describe
the crystal shape near the tip.  Accordingly, it may be more difficult
to use a power law fit to look for the onset of sidebranching or
possible tip oscillations.

Two remaining issues may be relevant for all of the models.  First,
the extent to which optical distortions near the tip affect the
results has not been addressed.  Specifically, since both the
concentration gradients and the interfacial curvature are largest near
the tip, the data closest to the tip are the least
reliable\cite{Glicksman95,Dougherty88}.  Second, the extent to which
all of these fits are contaminated by early sidebranches needs to be
investigated.  This may be especially important in characterizing the
emergence of sidebranches as well as in studies of possible tip
oscillations.




\bibliography{tipshape}
\bibliographystyle{elsart-num}

\end{document}